\documentclass[aps,pre,twocolumn,showpacs,superscriptaddress]{revtex4}
\usepackage{epsfig}
\usepackage{times}

\begin{document}

\title{Benefits of tolerance in public goods games}

\author{Attila Szolnoki}
\email{szolnoki@mfa.kfki.hu}
\affiliation{Institute of Technical Physics and Materials Science, Centre for Energy Research, Hungarian Academy of Sciences, P.O. Box 49, H-1525 Budapest, Hungary}

\author{Xiaojie Chen}
\email{xiaojiechen@uestc.edu.cn}
\affiliation{School of Mathematical Sciences, University of Electronic Science and Technology of China, Chengdu 611731, China}

\begin{abstract}
Leaving the joint enterprise when defection is unveiled is always a viable option to avoid being exploited. Although loner strategy helps the population not to be trapped into the tragedy of the commons state, it could offer only a modest income for non-participants. In this paper we demonstrate that showing some tolerance toward defectors could not only save cooperation in harsh environments, but in fact results in a surprisingly high average payoff for group members in public goods games.
Phase diagrams and the underlying spatial patterns reveal the high complexity of evolving states where cyclic dominant strategies or two-strategy alliances can characterize the final state of evolution. 
We identify microscopic mechanisms which are responsible for the superiority of global solutions containing tolerant players.
This phenomenon is robust and can be observed both in well-mixed and in structured populations highlighting the importance of tolerance in our everyday life.
\end{abstract}

\maketitle

\section{Introduction}

It is difficult to overestimate the importance of cooperation among players who are motivated to search for maximal individual income during their interactions with competitors \cite{pennisi_s05}. Although mutual cooperation would provide the optimal income for the whole community, a higher payoff can be reached individually by exploiting others. This conflict, summarized in several social dilemmas \cite{kollock_ars98}, can be identified as the key problem in a broad range of research fields \cite{axelrod_84,sigmund_10,nowak_11,frean_prsb13,yang_hx_njp14,matamalas_srep15}.

Staying with a specific example, it is always disappointing to realize when some of our partners defect in a working group, which significantly lowers the income of cooperator members. A natural reaction could be to punish the traitor, but the institution of punishment raises further questions, which sometimes just transfers the basic problem to another level \cite{fehr_n02,boyd_pnas03,helbing_ploscb10,rand_jtb09,shimao_pone13,li_k_pre15}. An alternative response from betrayed cooperators could be to stop further cooperation and not to participate in the joint venture anymore. Accordingly, cooperators may become ``loners'' because the latter strategy can offer a modest, but at least guaranteed, payoff to them. Previous works revealed that the option of voluntary participation in common ventures could be an effective way to avoid being exploited because it introduces a cyclic dominance between competing strategies of defectors, cooperators, and loners \cite{hauert_s02,szabo_prl02,semmann_n03}. As a consequence, the cooperator state can survive even in harsh conditions when a low synergy factor would result in a full defector state in a two-strategy system where participation in a public goods game is compulsory. There is, however, a disappointing feature of the new, three-strategy solution. Namely, the average payoff is unable to exceed the income of a loner's strategy, hence participating in a public goods game does not necessarily provide an attractive option for competing players \cite{hauert_s02,hauert_jtb02}.

This failure suggests that perhaps it is not the best option for cooperators to leave the group when defectors emerge because by switching to a loner state they lose all benefits of mutual cooperation immediately. In this way the original dilemma can be transformed into a new form where cooperator players should decide how many defectors they tolerate in their group before leaving the group for a modest, but guaranteed payoff.
  
To explore this new dilemma we introduce a four-strategy model of a public goods game in which besides the unconditional defector ($D$), cooperator ($C$), and loner ($L$) strategies there is a so-called tolerant or mixed ($M$) strategy, that behaves as a cooperator as long as the number of defectors remains below a threshold value in the group but it switches to loner a state otherwise. By following this approach we can check the viability of this mixed strategy and clarify if there is an optimal level of tolerance which provides the highest income for the whole population.

Beyond these fundamental questions there is an additional aspect which makes the proposed model even more interesting. On one hand, the coexistence of $C$, $D$, and $L$ strategies is based on the previously mentioned cyclic dominance between competing strategies, which is a well identified general mechanism to maintain diversity \cite{kerr_n02,kirkup_n04,reichenbach_pre06,arenas_jtb11,wang_wx_pre11,szolnoki_jrsif14,groselj_pre15}. On the other hand, by considering $M$ players we introduce a strategy which is less harmful to defectors because they may coexist. Intuitively, one may expect that such intervention is beneficial to defection, but, as we demonstrate in this paper, the opposite effect can be observed.

The organization of this paper is as follows. We present the definition of the model in the next section. Results obtained by means of the replicator equation in well-mixed populations are summarized in Section~\ref{results}, which is followed by the presentation of Monte Carlo results obtained in structured populations. Finally we conclude with an argument for broader validity of our observations and a discussion of their implications in Section~\ref{summary}.

\section{Public goods game with tolerant players}
\label{def}

We consider a public goods game where the game is played in groups of size $G$. Following the standard model \cite{hauert_s02}, each player is set as an unconditional cooperator ($C$), an unconditional defector ($D$), or a loner ($L$). Whereas each cooperator contributes an amount $c$ to the common pool, defectors contribute nothing but exploit others' efforts. Loners do not participate in the joint enterprise, instead, they prefer a moderate, but guaranteed, $\sigma$ income. Beyond these well-known strategies we consider an additional, so-called mixed ($M$) strategy. The latter players are principally cooperators who contribute to the common pool but permanently monitor the status of other players in the group at an additional cost of $\gamma$. This extra knowledge allows them to realize if the level of defection exceeds a certain level in the group. As a reaction, they become loners and stop contributing to the common pool.  
Designating then the number of unconditional cooperators, defectors, and ``mixed'' players among the other $G-1$ players in the group as $n_C, n_D$ and $n_M$, the payoff of the four competing strategies are the following:
\begin{eqnarray}
\Pi_D\,\, &=& \frac{r (n_C+\delta n_M)}{n_C+n_D+1+\delta n_M}\\
\Pi_C\,\, &=& \frac{r (n_C+1+\delta n_M)}{n_C+n_D+1+\delta n_M} - 1\\
\Pi_L\,\, &=& \sigma\\
\Pi_M &=& \delta \left[\frac{r (n_C+1+n_M)}{n_C+n_D+1+n_M} - 1\right] + (1-\delta) \sigma - \gamma\,.\,\,\,\,\,\,\,\,\,
\label{payoff}
\end{eqnarray}
Here, according to the broadly accepted notation, $r$ depicts the synergy factor, characterizing the benefit of mutual cooperation, whereas $\sigma$ is the loner's payoff. It should be emphasized that the $r>1$ synergy factor is applied only if there are more than one contributor to the common pool, otherwise $r=1$ is used. In this way we can avoid an artificial support of a lonely cooperator against loners and prevent single individuals playing a public goods game with themselves \cite{brandt_pnas06}. Furthermore, without loss of generality, cooperators' contribution to the common pool is considered to be $c=1$, as Eqs.~(2) and (4) indicate. Lastly, in close agreement with previous works \cite{hauert_s02,hauert_jtb02}, the payoff of loners is chosen as $\sigma=1$, but we stress that using other values would not change our main findings.

As we already noted, it is a fundamental point that an $M$ player uses a more sophisticated strategy by checking the status of other players in the group. Accordingly, such a player behaves as a loner and refuses participating in the public goods game if the number of defectors reaches a critical $H$ threshold in the group. Otherwise, when the total number of defectors is below the $H$ threshold, $M$ cooperates and contributes to the common pool similarly to unconditional cooperators. The possible ``switch'' of a player's status, or saying differently the adoption to change a neighborhood, can be handled technically via a $\delta$ factor, which is $\delta=0$ if $n_D \ge H$ or $\delta=1$ if $n_D < H$. 

Formally, strategy $L$ can be considered as a cost-free, very special mixed player who applies zero threshold, hence he always avoids participating in a joint venture independent of the strategies of other group members. 
In general, the value of $H$ characterizes the level of tolerance of $M$ players. Namely, the higher $H$ is applied, the more defectors are accepted in the group without refusing cooperation from $M$ players. As an extreme case, formally $H=G$ denotes the situation when $M$ players remain in an unconditional cooperator state. Hence we may say that the concept of ``tolerance'' builds a bridge between loner and unconditionally cooperator behaviors.
Close to the latter end, $H=G-1$ represents the case when an $M$ player seems to be almost ``endlessly tolerant'' and becomes a loner only if all the others in the group are defectors, hence cooperation becomes unambiguously pointless to him.

Evidently, the extra knowledge of $M$ players needs additional efforts from their side, which can be implemented via an additional cost $\gamma$. This cost should always be considered, no matter whether $M$ plays a $C$ or $L$ strategy, as indicated in Eq.~(4). 
The presence of this permanent cost also means that $M$ players have no obvious advantage either over $C$ or over $L$ strategies.

In the following we consider both well-mixed and structured populations. 

\section{Results}
\label{results}

\subsection{Well-mixed populations}

In a well-mixed system the fraction of $C, D, L$, and $M$ players can be denoted by $x, y, z$, and $w$ respectively. Evidently, they are not independent but are normalized and  always fulfill the equation $x+y+z+w=1$. Consequently, the strategy evolution can be studied by using replicator dynamics \cite{hofbauer_98}:
\begin{eqnarray}
\dot{x} &=& x (P_C - \overline{P}) \nonumber\\
\dot{y} &=& y (P_D - \overline{P})\\
\dot{z} &=& z (P_L - \overline{P}) \nonumber\,\,,
\label{replicator}
\end{eqnarray}
where dots denote the derivatives with respect to time $t$. Here the average payoff $\overline{P}$ for the whole population is given by
\begin{equation}
\overline{P} = x P_C + y P_D +z P_L + w P_M \,,
\label{averP}
\end{equation}
where $P_i$ ($i={C,D,L,M}$) designates the average payoff for each strategy:
\begin{equation}
P_i = \sum_{n_C,n_D,n_L,n_M} \frac{(G-1)!}{n_C!n_D!n_L!n_M!} x^{n_C} y^{n_D} z^{n_L} w^{n_M} \Pi_i
\end{equation}  
where $0\le n_i \le G-1$ and $\sum n_i = G-1$ are always fulfilled.

For the sake of comparison with the case of a structured population we suppose that players form groups of size $G=5$ randomly and consider the impact of different $H=1,\dots,G-1$ threshold values of tolerance.

\begin{figure}
\centerline{\epsfig{file=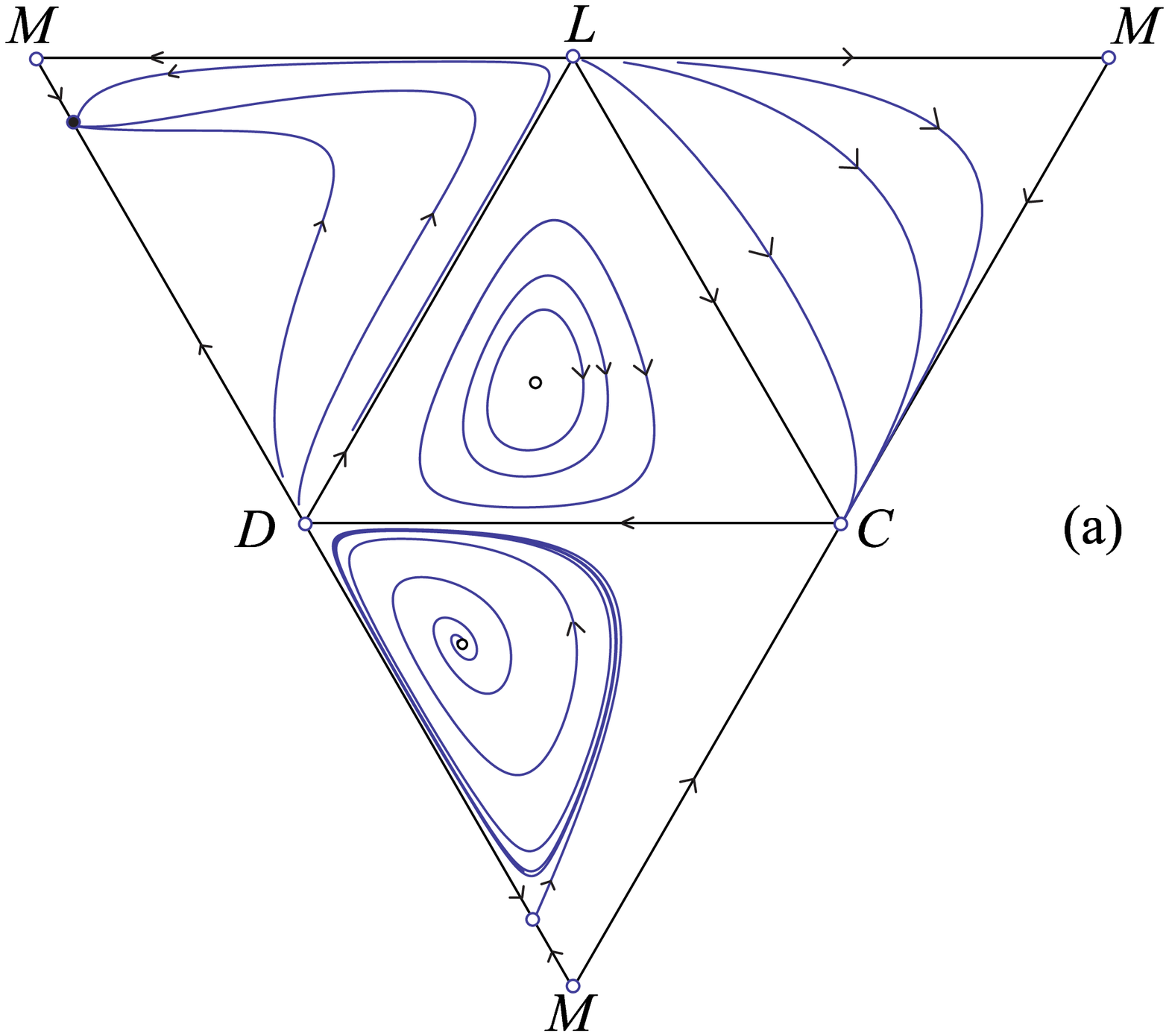,width=6.2cm}}
\centerline{\epsfig{file=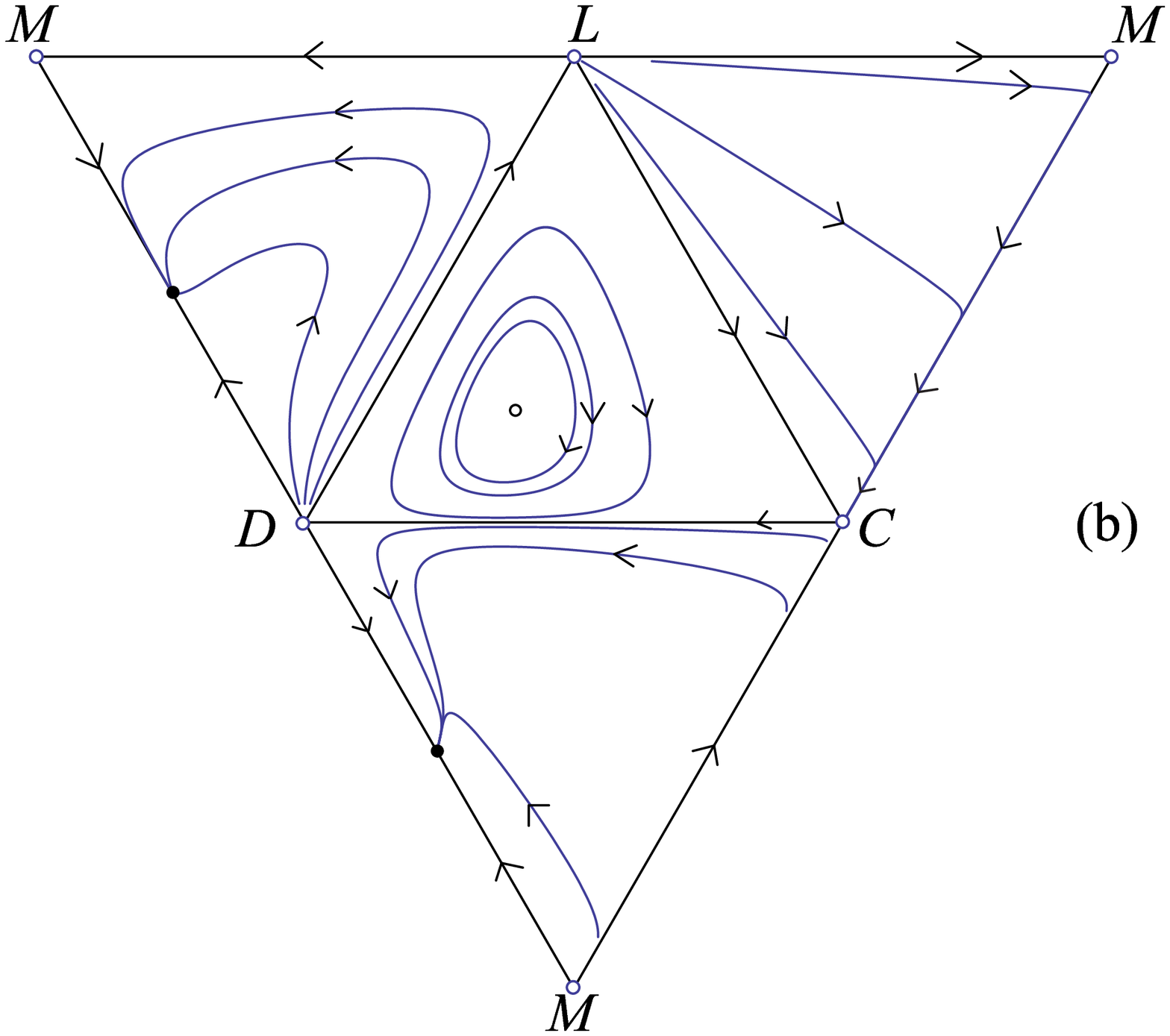,width=6.2cm}}
\caption{\label{simplex} (Color online) Replicator dynamics on the boundary faces of the simplex $S_4$ using the $G=5$ group size. Filled circles represent stable fixed points whereas open circles represent unstable fixed points. Parameter values are $H=2, r=3.5, \gamma=0.6$ for panel~(a), whereas $H=4, r=3.8, \gamma=0.04$ for panel~(b). Flow diagrams suggest that both $(D,C,L)$ and $(D,C,M)$ strategies can form a rock-paper-scissors cycle, but the stable two-strategy ($D,M$) phase also emerges in dependence on the initial fraction of strategies.}
\end{figure}

Our principal goal is to compare the results of a well-mixed and spatially structured population, therefore we will launch the evolution from a random initial state where all competing strategies are present with equal weight. The replicator dynamics, however, may depend sensitively on the initial frequencies of strategies. This behavior is illustrated in Fig.~\ref{simplex} where we have plotted two representative flow diagrams in the unit simplex $S_4$ at two branches of parameter values. 

The top panel illustrates the case when the $H=2$ threshold value is applied at a significantly high $\gamma$ cost of inspection when the synergy factor is moderate. Here, as is already known from a previous work \cite{hauert_s02}, the $w=0$ face contains a fix point which is surrounded by periodic orbits. On the $z=0$ face, however, there is a stable limit circle which is the composition of $(D,C,M)$ strategies. Furthermore, a stable two-strategy fix point can also be detected on the $x=0$ face. In the bottom panel, which was taken at the $H=4$ threshold level, we can observe that the $(D,M)$ solution remains stable whereas the rock-scissors-paper-type $(D,C,M)$ solution disappears. Naturally, the portrayal of replicator dynamics can also depend on the applied $r$ and $\gamma$ parameters, but the presented plots are representative in a broad interval of parameters.
 
In the following we focus on the evolution from a random initial state where all strategies are present with equal weight, but we scan the whole $r - \gamma$ parameter plane.  
Interestingly, when $H=1$ then strategy $M$ cannot survive at any finite values of $\gamma$. Here, $(D,C,L)$ strategies form the well-known rock-scissors-paper type solution in the $2 < r < 5$ region \cite{hauert_s02}. At low $\gamma$ values, however, $M$ players may crowd out loners first from a random state, which is followed by the extinction of defectors. Finally, when $M$ remains alone with unconditional $C$ players, $M$ is defeated by the latter strategy. This time evolution is similar to the ``The Moor has done his duty, the Moor may go'' effect previously observed in a related model where punishing strategies were studied \cite{szolnoki_pre11b}. Nevertheless, we should emphasize that the only stable solution is the mentioned three-strategy $(D,C,L)$ state at $H=1$.

\begin{figure}
\centerline{\epsfig{file=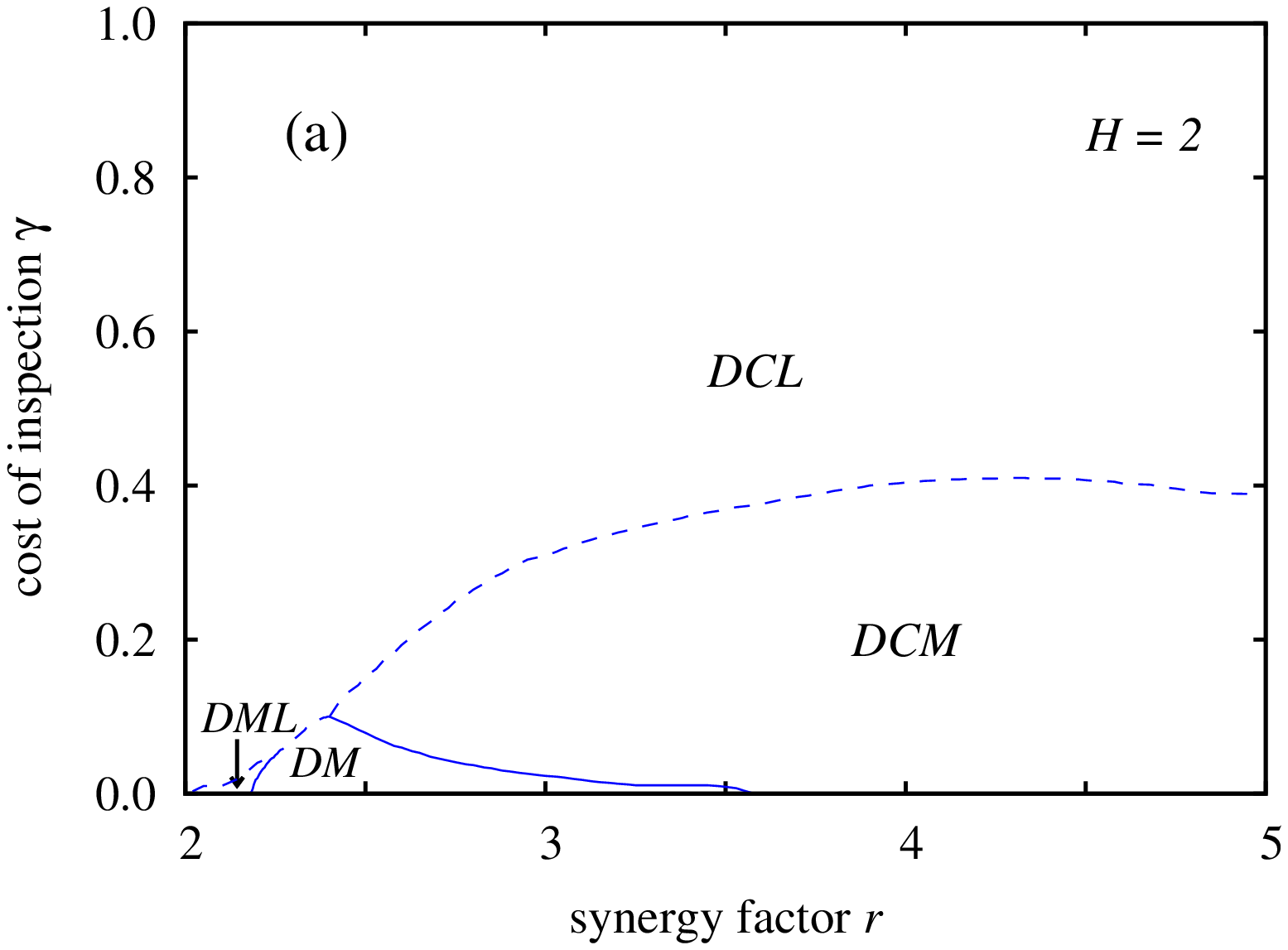,width=8.0cm}}
\centerline{\epsfig{file=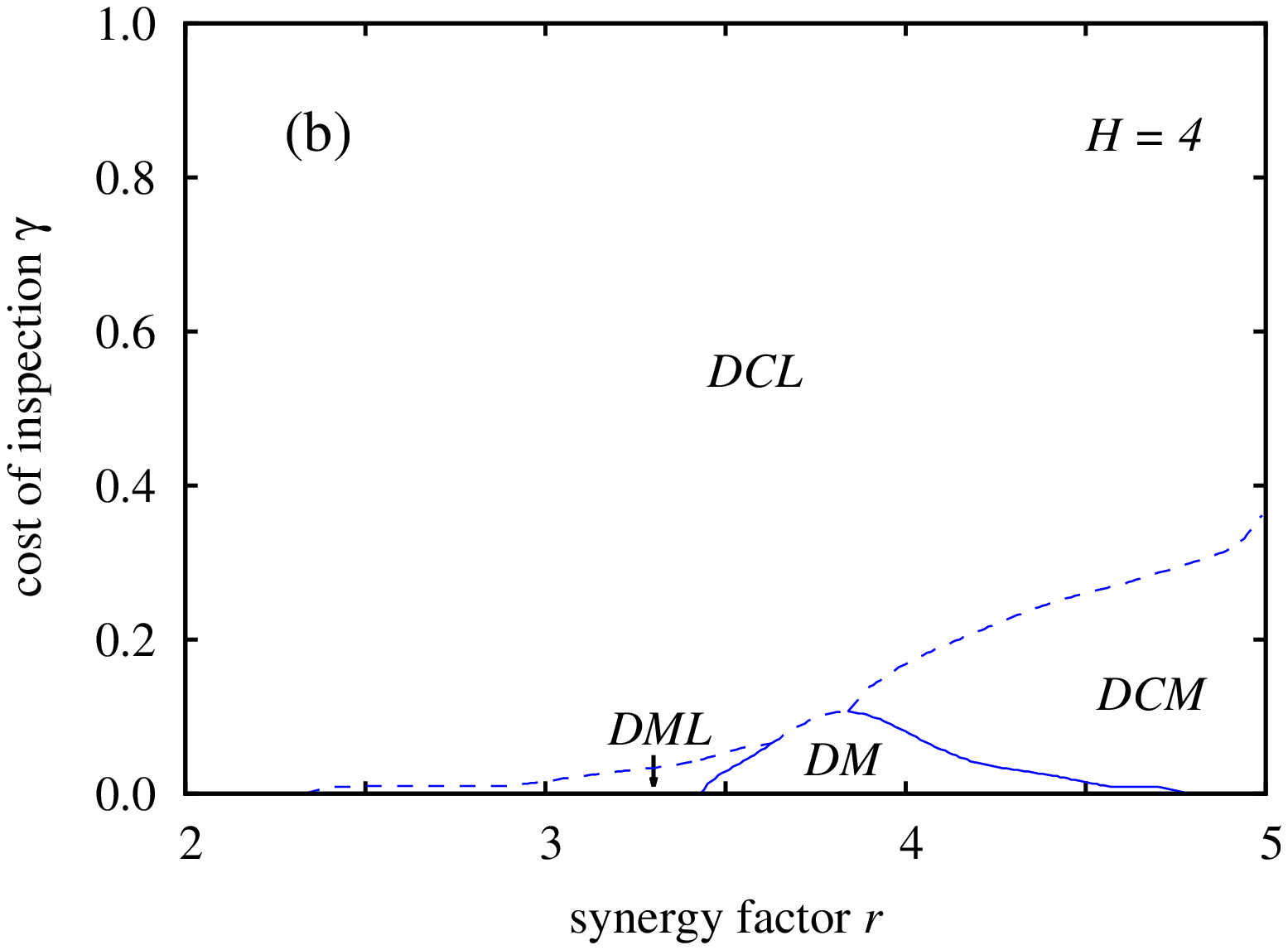,width=8.0cm}}
\caption{\label{mf}  (Color online) Full $r-\gamma$ phase diagrams for the well-mixed system in the case of $G=5$ group size. In panel~(a)
$H=2$ whereas in panel~(b) $H=4$ threshold values are applied. 
Solid lines represent continuous, while dashed lines indicate discontinuous transitions between stable solutions. At moderate $\gamma$ values $M$ players replace loners by forming a cyclic dominant coexistence with $C$ and $D$ players. $M$ players are more viable at an intermediate threshold value. Interestingly, strategy $M$ can form a two-strategy alliance with $D$ which can dominate the evolution at specific parameter values.}
\end{figure}

By increasing the tolerance level, however, we can observe new types of solutions. Namely, strategy $M$ can replace $L$ players and forms another solution where $D$, $C$, and $M$ players dominate each other cyclically. As the top panel of Fig.~\ref{mf} illustrates, this $(D,C,M)$ state can be dominant even at a significantly high $\gamma$ cost if synergy factor $r$ is high enough. The latter condition, when mutual cooperation pays more, is essential, otherwise the benefit of mutual cooperation could not compensate the additional cost of strategy $M$.

The previously mentioned two-strategy solution can evolve starting from a complete, four-strategy initial state. Here $D$ and $M$ players form a two-strategy alliance against other competing strategies \cite{szabo_pre01b,perc_pre07b}. Note that unconditional cooperators would beat strategy $M$ in the absence of defectors, but the presence of latter players manifests the advantage of a mixed strategy. This solution, as we emphasize in the subsequent section, is of prime importance to understand why tolerance emerges during an evolutionary process. Lastly, we briefly note that there is a specific combination of $D, L,$ and $M$ players which could prevail in the whole system, albeit at a very limited parameter region. 

Qualitatively similar behavior can be observed for other $H>1$ threshold values, too, but solutions containing mixed players become less vital as we increase $H$. The bottom panel of Fig.~\ref{mf}, obtained at $H=4$, illustrates that the benefit of mixed strategy is less likely at such a high tolerance level and the area where strategy $M$ survives on the $\gamma-r$ parameter plane shrinks significantly.

\begin{figure}
\centerline{\epsfig{file=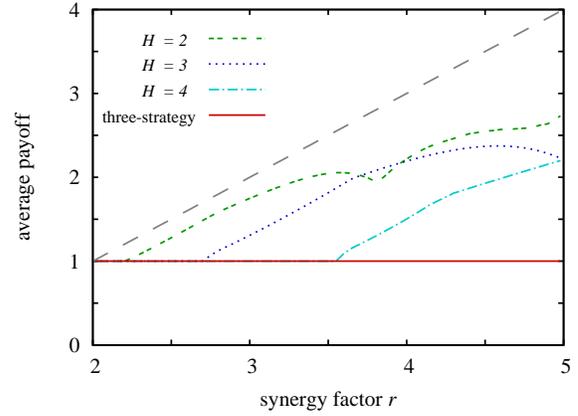,width=8.0cm}}
\caption{\label{mf_payoff}  (Color online) Average payoffs in stable stationary states in dependence of synergy factor $r$ as obtained for different threshold values of tolerance at $\gamma=0.04$. Players are in a well-mixed population where they form groups of size $G=5$ randomly. For comparison, we also show the maximum reachable average payoff that can be obtained in the state, marked by a dashed grey line, where all players are in unconditional cooperator states. Note that in the case of $H=1$ the stable solution is the traditional three-strategy state when $D, C,$ and $L$ players form a cyclic dominant solution \cite{hauert_s02}. Here the average payoff cannot exceed the loner's $\sigma=1$ income. By using a bit higher tolerance level, however, a significant improvement can be reached, which is comparable to the value obtained in an idealistic (all $C$) state.}
\end{figure}

It is an important consequence that the introduction of tolerance does not only result in the individual success of strategy $M$, but also has a favorable impact on the general well-being of the whole population. This effect can be illustrated nicely if we compare the average payoff values obtained at different threshold levels. Figure~\ref{mf_payoff} highlights that adopting a nonzero, but moderate tolerance towards defection could elevate significantly the global income of players. As we already noted, 
the usage of the minimal $H=1$ tolerance level does not allow $M$ players to survive, hence the system becomes equivalent to the well-known three-strategy model \cite{hauert_s02}. Here the presence of $L$ players can help to avoid the tragedy of the common state \cite{hardin_g_s68}, but the average payoff cannot exceed the loners' income that is $\sigma=1$ in the present case. At a higher tolerance threshold,
the average income in the whole population can increase significantly due to the presence of tolerant players. This enhancement is particularly conspicuous for $H=2$. For comparison, the dashed grey line shows the average income in the idealistic state when all players are in an unconditional cooperator state. It suggests that the usage of tolerant players could be especially efficient at low $r$ values when cooperators would face a harsh environment otherwise.

\subsection{Structured populations}

Considering a structured population where players have fixed neighborhood offers not just a more appropriate approach to some real-life situations but it often provides novel and sometimes unexpected behaviors which are absent in a well-mixed system \cite{nowak_n92b,santos_prl05,szabo_pr07}. To explore and clarify the possible differences between emerging solutions we consider a structured population where players are distributed on a square graph and form groups with their nearest neighbors ($G=5$). It also means that a player is involved not only in the game where it is a focal player but also in the games of his neighbors. Therefore a player may participate in $G=5$ public goods games, and the total payoff should be accumulated accordingly. 

During the strategy update protocol, we apply strategy imitation based on pairwise comparison of competing strategies \cite{szabo_pr07}. Namely, a player $x$ will adopt the strategy of a neighboring player $y$ with a probability
\begin{equation}
\Gamma (\Pi_x-\Pi_y)= \frac{1}{1+\exp((\Pi_x-\Pi_y)/K)}\,\,,
\label{fermi}
\end{equation}
where $K$ is the noise parameter. Without loss of generality we will use a representative $K=0.5$ value, which ensures that strategies of better-performing players are  adopted almost always by their neighbors, although adopting the strategy of a player that performs worse is not impossible.

In an elementary step, we choose a player and his neighbor randomly. If their strategies are different then the strategy imitation is executed with the probability defined by Eq.~\ref{fermi}. In a complete Monte Carlo step (MCS) every player has one chance on average to update his strategy.
To get reliable phase diagrams, which are valid in the large system size limit, the system size was chosen from $400\times 400$ to $6400 \times 6400$, and the relaxation time was between 20000 and 100000 MCS. 
To further improve accuracy, the results of the stationary state were averaged over 10 independent realizations for each set of parameter values.

\begin{figure*}
\centerline{\epsfig{file=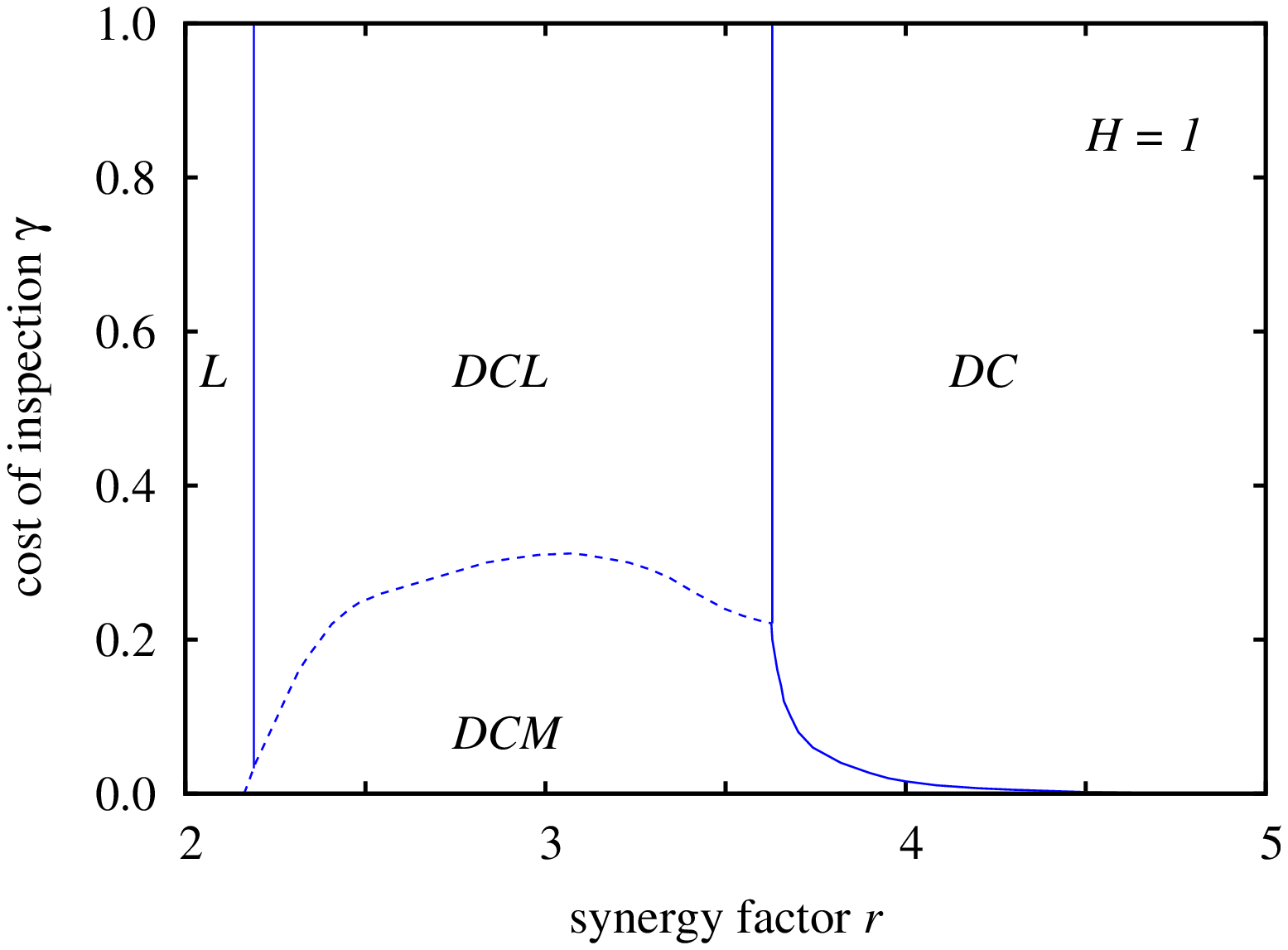,width=8.0cm}\epsfig{file=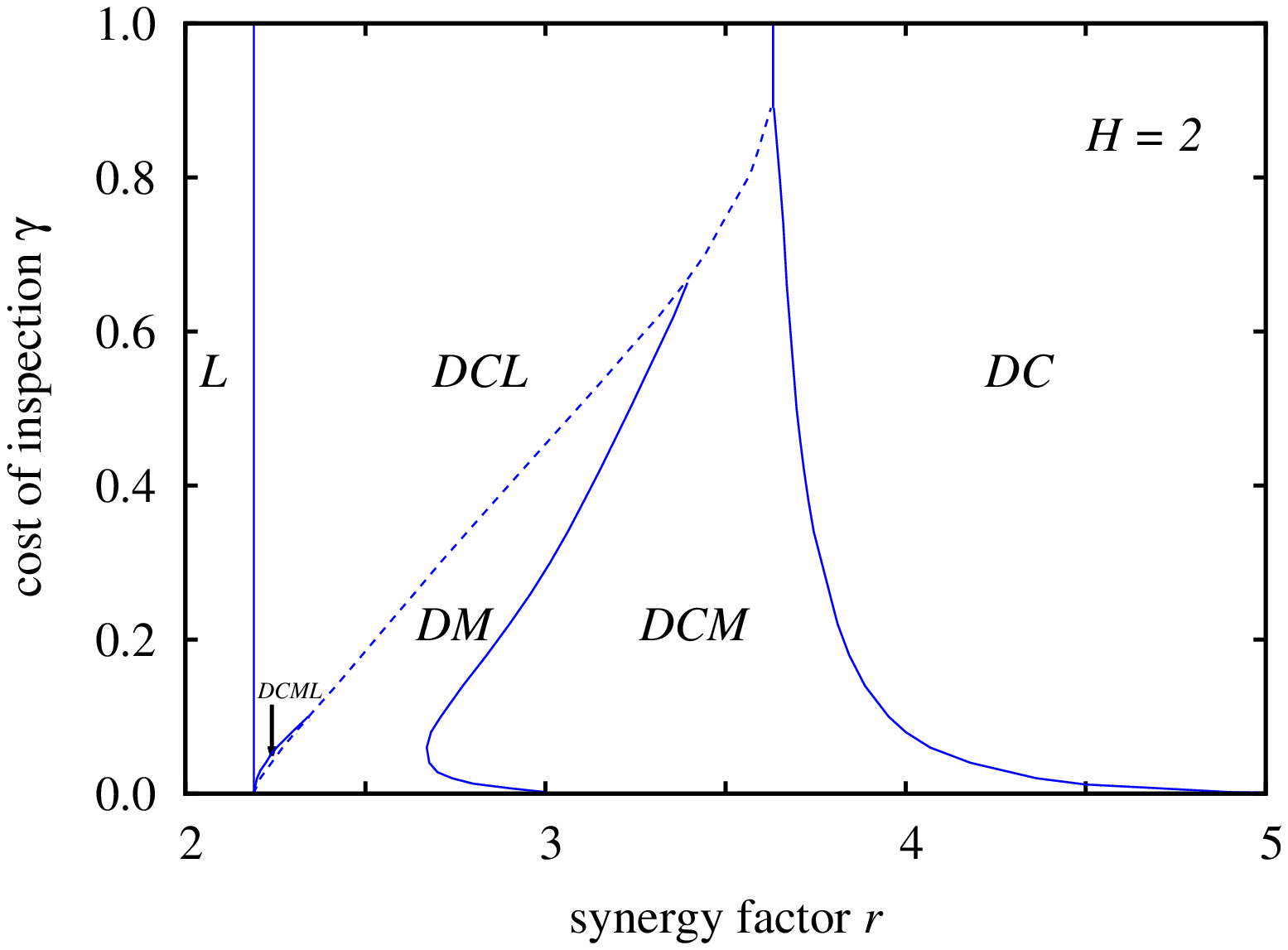,width=8.0cm}}
\centerline{\epsfig{file=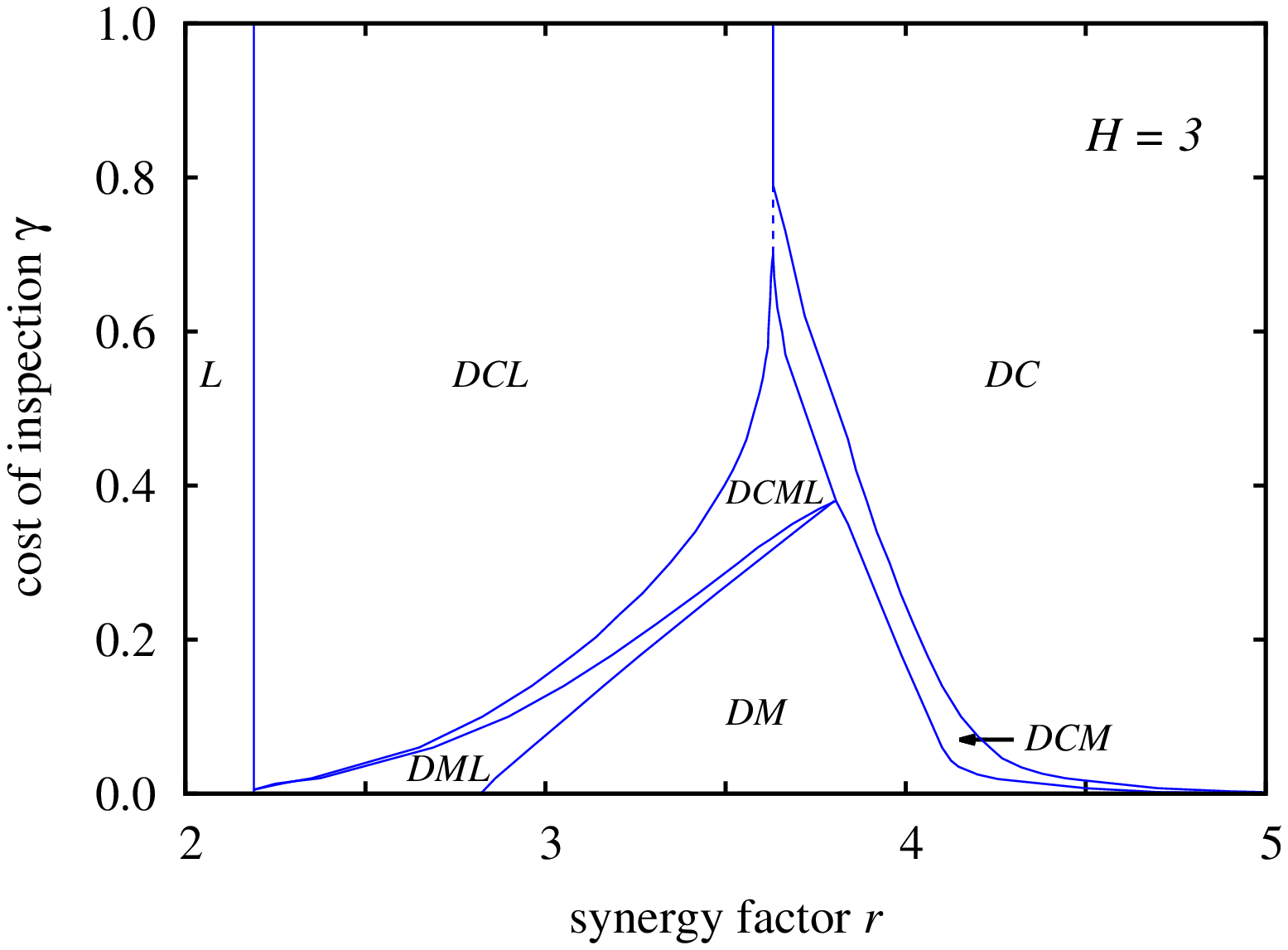,width=8.0cm}\epsfig{file=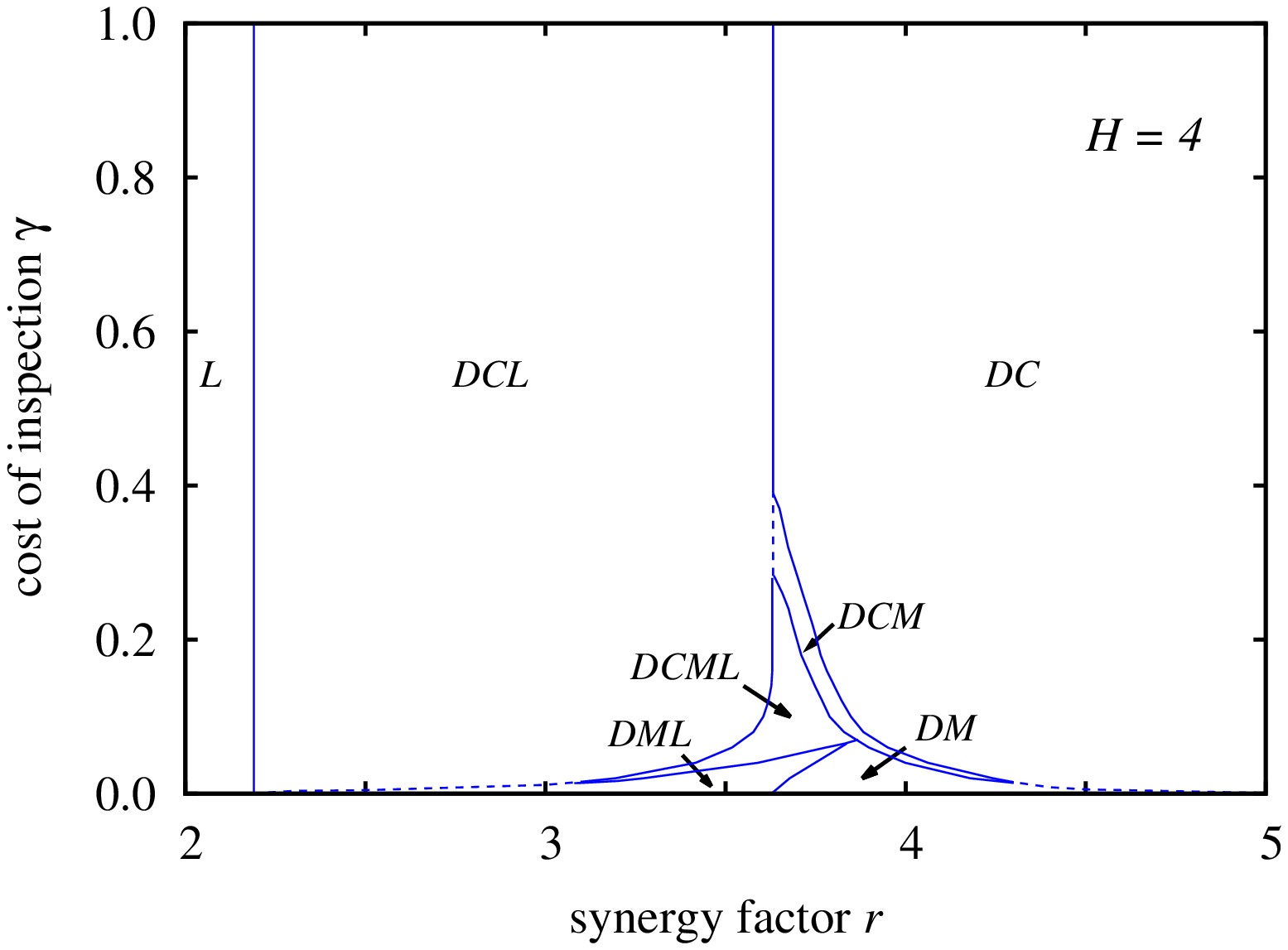,width=8.0cm}}
\caption{\label{phd}  (Color online) Full $r-\gamma$ phase diagrams for the spatial public goods game where players are distributed on a square lattice forming $G=5$ size of groups. Different $H$ threshold values are indicated. Solid lines represent continuous, whereas dashed lines indicate discontinuous transitions between stable solutions. The comparison of diagrams shows that a moderate tolerance, an intermediate threshold value of $H$, allows $M$ players to prevail even at a significantly high cost value.}
\end{figure*}

We should stress that the evolution of strategies in a structured population is highly independent of the initial state if all competing strategies are present. The only critical condition is the sufficiently large system size which prevents finite-size effects and allows a stable solution to emerge somewhere in a space from a random initial distribution. Later this solution can invade the whole space and remains stable. 
 
In our model there are three key parameters, namely the $r$, $\gamma$, and $H$ threshold level. To demonstrate their impacts on the stable solutions we have presented the resulting phase diagrams for the four possible threshold values in Fig.~\ref{phd}.

In general, as expected, strategy $M$ always dies out if the inspection cost is too large and we get back the well-known three-strategy $(D,C,L)$ model. In this case unconditional cooperators can survive above $r \ge 2.19$ due to cyclic dominance and form a three-strategy ($D,C,L$) phase. If the synergy factor is high enough and strategy $C$ is capable to coexist with $D$ due to network reciprocity then the previously mentioned cyclic dominance is broken, which will result in the extinction of strategy $L$. Consequently, a two-strategy ($D,C$) phase remains where the fraction of defectors decreases gradually as we increase $r$. 

Significantly different behavior can be obtained if the extra $\gamma$ cost of $M$ players is reasonably moderate. As a general observation, which is partly against mean-field results, 
strategy $M$ becomes viable, but the composition of the stable solution depends sensitively on the threshold value of tolerance.
The lowest nonzero $H=1$ value represents a special case because here $M$ can only survive with $D$ in the presence of $C$ players. Due to this low threshold an $M$ player changes from the $C$ to the $L$ state immediately when it recognizes the presence of a defector in the group, hence the previous mentioned cyclic dominance is established, but instead of the $(D,C,L)$ cycle the strategies $M \to D \to C \to M$ will form the stable three-strategy solution. In other words, $L$ is simply replaced by $M$ who takes the role of the former strategy. The advantage of a three-strategy solution over the other state depends on the average rotation speed between cyclic members: If the invasion rate is faster, then it can stabilize a solution \cite{perc_pre07b}. By increasing $r$ we may observe a reentrant transition between $(D,C,L) \to (D,C,M) \to (D,C,L)$ phases, which is again a general behavior when the average invasion rates within a cycle can be adjusted by varying a control parameter \cite{szolnoki_epl15}.  

If we increase $H$ and allow $M$ players to ``tolerate'' the presence of defectors further then a new kind of solution emerges, which was already observed in the well-mixed system. In this case $M$ can coexist with $D$ without the presence of a third party. As we will show later, this $(D,M)$ solution can be specially efficient to reach a state when a high average payoff can be reached for the whole population. Besides the mentioned two-strategy solution, there are parameter values where all four competing strategies coexist, and there are some specific cases when $M$ crowds out unconditional $C$ but stay together with $L$ in the presence of $D$. Here $D$ and $M$ players are still capable of forming a two-strategy solution, but $L$ players can invade defectors. As a result, small $L$ patches emerge temporarily, but they are vulnerable against the invasion of $M$ players, who are capable of utilizing network reciprocity, which closes the cycle. 

Figure~\ref{phd} highlights that the new kind of solution can also emerge in a structured population. In particular, we can observe a stable coexistence of four strategies (marked by $DCML$ in phase diagrams) which is absent in a well-mixed system. Such a kind of coexistence of competing strategies is a general feature of structured populations which is a straightforward consequence of the limited interactions of players \cite{nowak_n92b}.

The comparison of phase diagrams obtained for different tolerance thresholds highlights that there is an optimal intermediate tolerance level which provides the best condition for $M$ players. In this case strategy $M$ can survive even at a significantly high inspection cost. Note that an $M$ player should always bear this cost but has to invest also in the common pool when it cooperates. At $H=2$, for instance, $M$ should pay nearly double cost of the unconditional $C$ strategy, still, it can crowd out both $C$ and $L$ strategies. On the other hand, such a high ``peak'' is missing both at $H=1$ and at $H=4$, which can be considered as extreme (too high or too low) threshold values. 

Based on the comparison of phase diagrams we can conclude that neither too small nor too high tolerance will help $M$ players survive and they become extinct at relatively small $\gamma$ values. This observation agrees with our previous experiences obtained for a well-mixed population. It is worth stressing, however, that a tolerant strategy prevails more easily in a structured population and $M$ players can survive even at extremely high additional cost $\gamma$. 

\begin{figure}
\centerline{\epsfig{file=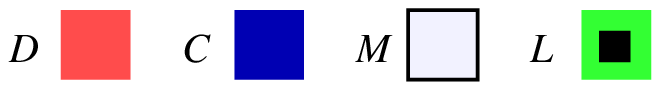,width=3.5cm}}
\centerline{\epsfig{file=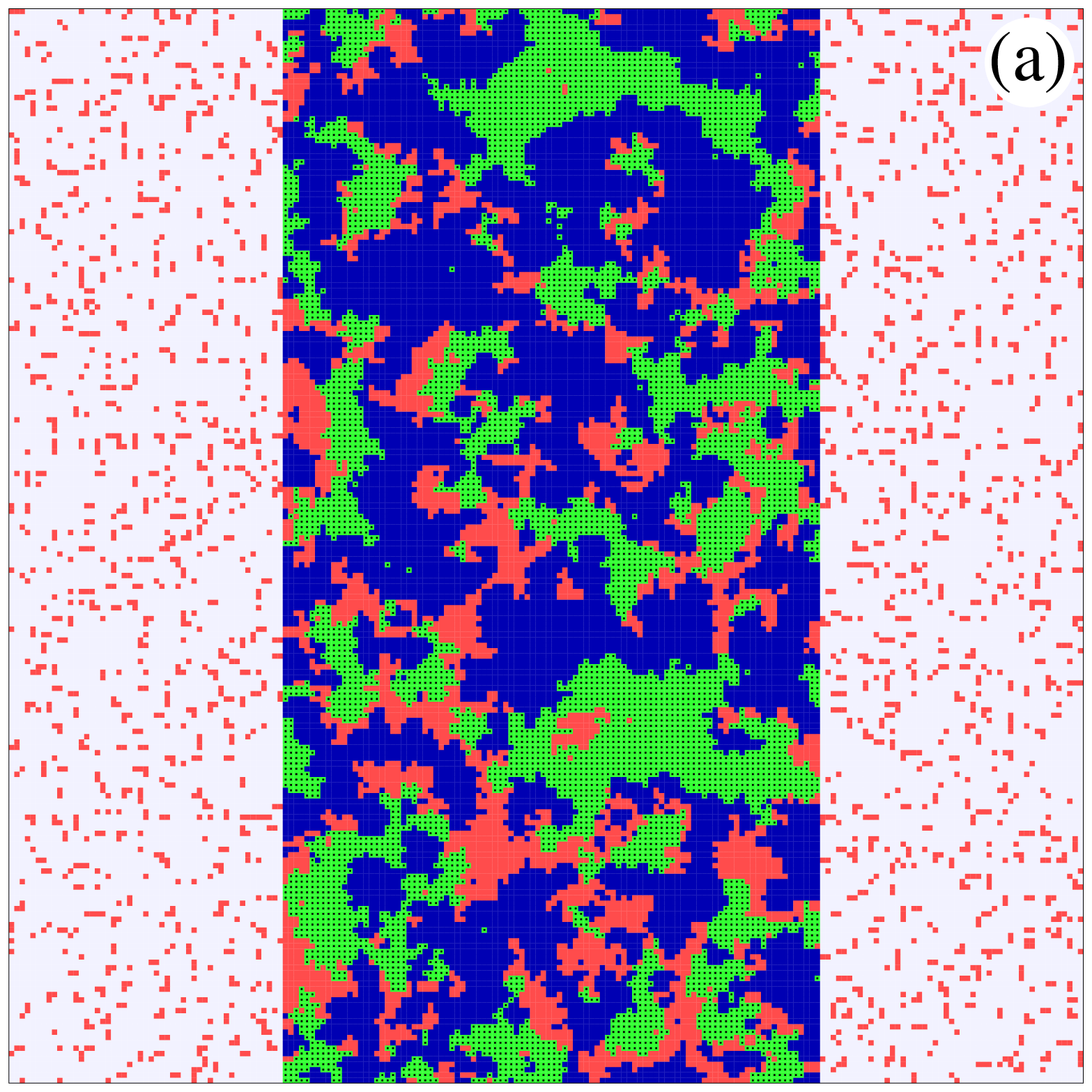,width=2.8cm}\epsfig{file=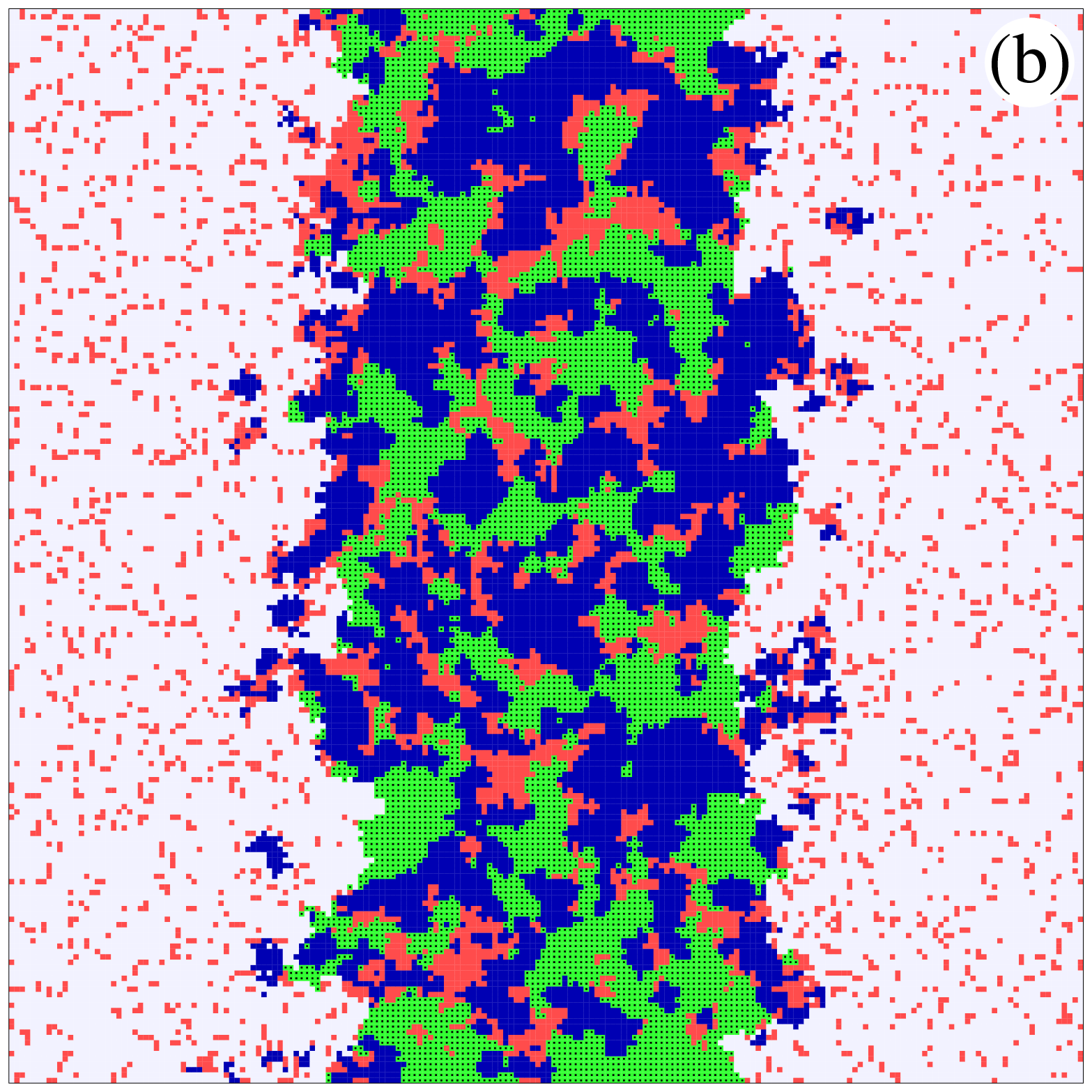,width=2.8cm}\epsfig{file=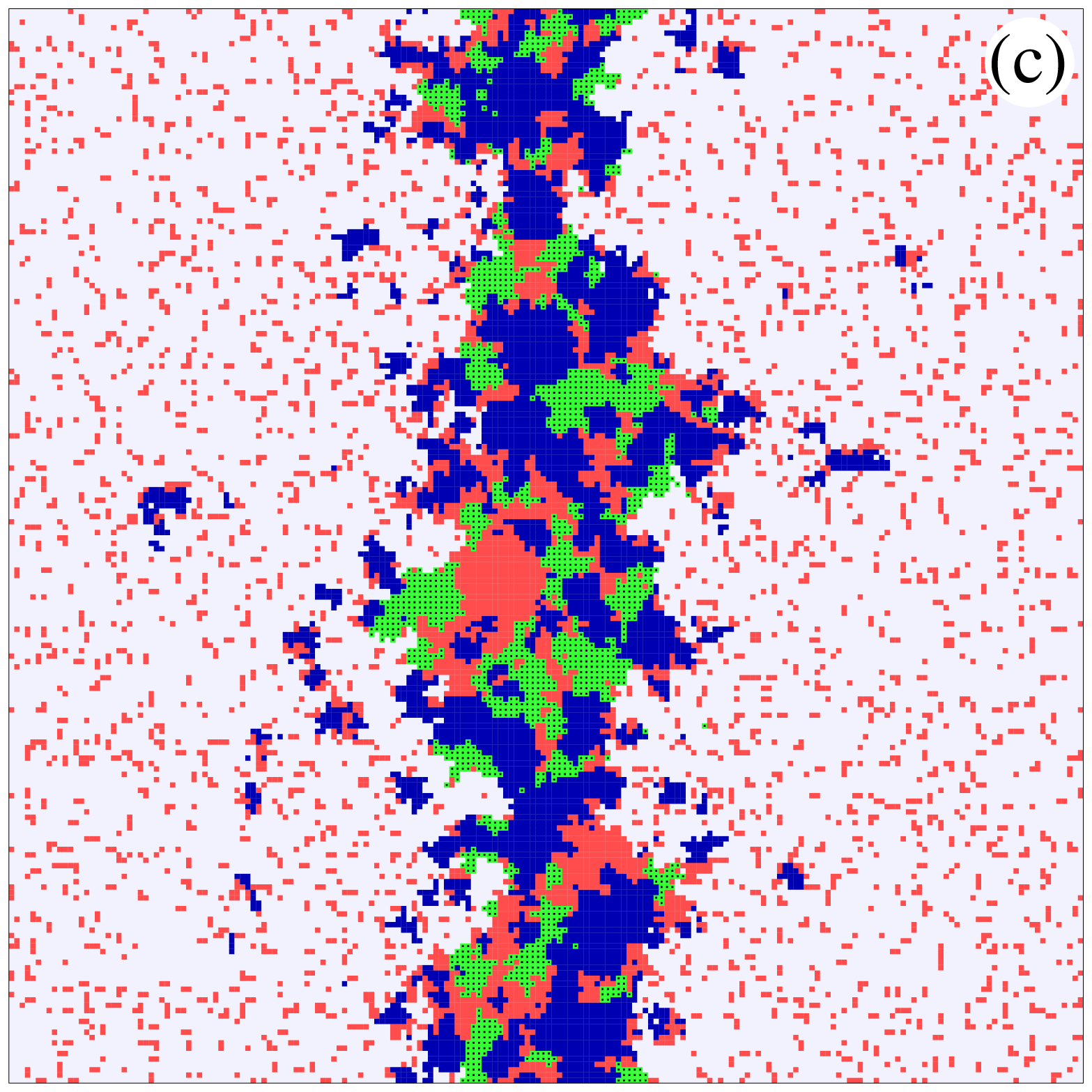,width=2.8cm}}
\centerline{\epsfig{file=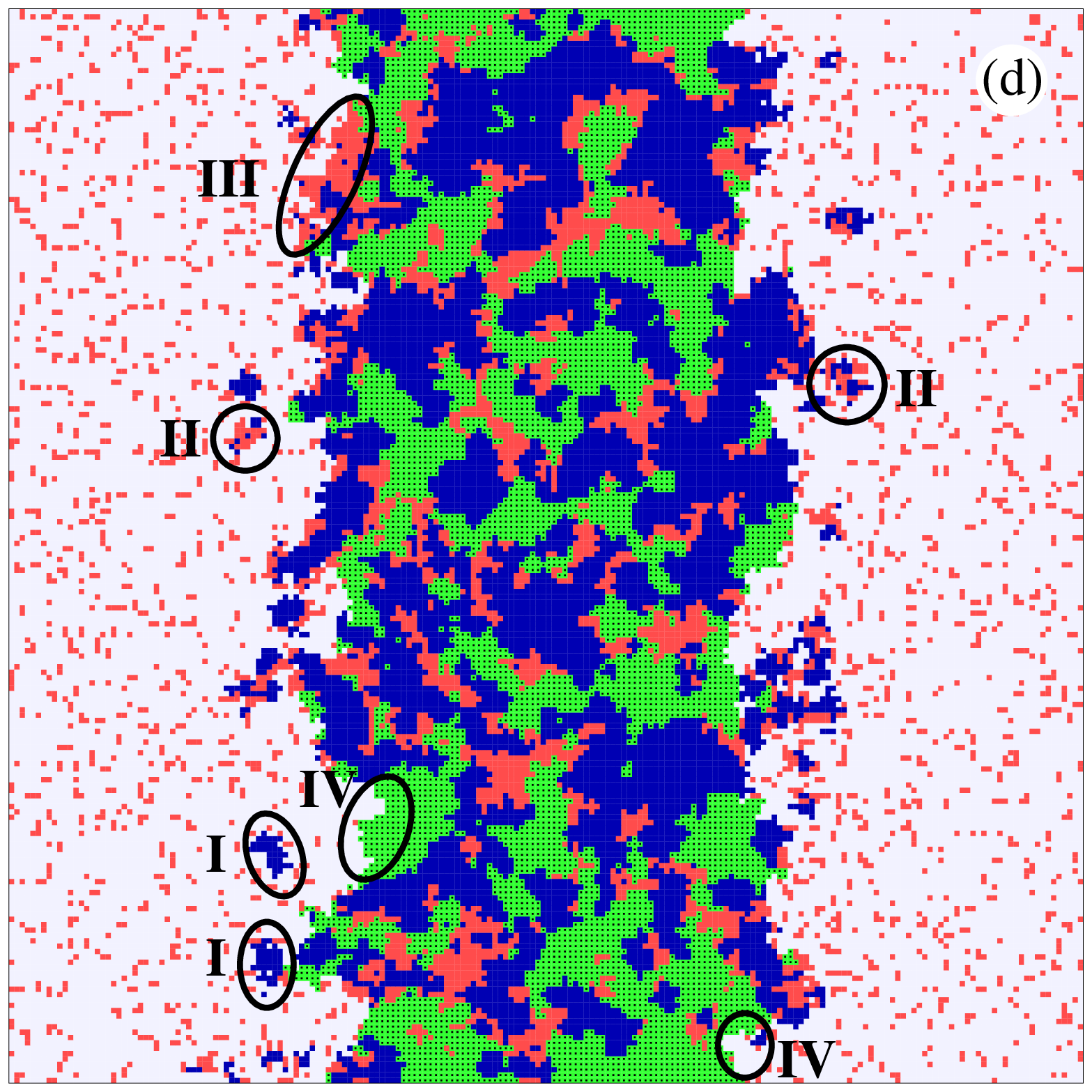,width=8.5cm}}
\caption{\label{tolerance_works} (Color online) The competition of two possible solutions at $r=2.7, \gamma=0.15$, using $H=2$ on a $200 \times 200$ square lattice. Defectors are denoted by red (middle grey), unconditional cooperators are denoted by dark blue (dark grey), tolerant players are denoted by light blue (light grey), whereas loners are denoted by a green color (dotted lighted grey), as indicated by the legend on the top. At the specific parameter values both the cyclic dominant $(D,C,L)$ and two-strategy $(D,M)$ phases could be possible solutions. Panel~(a) illustrates a prepared initial state where a cyclic dominant solution is embraced by the other stable solution. In panel~(b) we opened the borders and allowed solutions to compete for space. Eventually the $(D,M)$ solution crowds out the other phase, as is illustrated in panel~(c), and finally the two-strategy phase prevails (not shown). 
Panel~(d) shows the enlarged part of panel~(b) to illustrate the microscopic mechanisms that are responsible for the successful invasion of the $(D,M)$ solution.  
Further details are given in the main text. 
Snapshots were taken at 0, 70, 210 $MCS$s. }
\end{figure}

In the following we provide an intuitive explanation why tolerance can offer a viable way to handle defection.
It should be emphasized that the three-strategy $(D,C,L)$ phase is always a solution in the low-$r$ region \cite{hauert_s02}. To understand the superiority of the $(D,M)$ phase we will start the evolution from a special, prepared initial state where both the cyclically dominated phase and the stable coexistence of $D$ and $M$ players could evolve calmly in a restricted area first. Panel~(a) of Fig.~\ref{tolerance_works} illustrates the final result of these isolated evolutions. After, we let the borders open, and the battle of solutions starts. The elementary steps of this competition are identified in panel~(b), which is zoomed out for clarity. In this snapshot we can distinguish three different cases of how the three-strategy solution meets with the external two-strategy $(D,M)$ phase. If a $C$ domain, marked by dark blue, is at the frontier then unconditional cooperators start spreading in the sea of $M$. [These invasions are marked by ``I'' in panel~(b).] The success of $C$, however, is temporary, because defectors, marked by red, will follow them and gradually invade the invaders. [This stage is marked by ``II'' in panel~(b).] After, when $D$ players remain alone with $M$ players then the latter (marked by light blue) will regulate defectors and lower their concentration to a minimal level. The second option of how competing solutions meet is when a $D$ spot from the $(D,C,L)$ phase meets with the external $(D,M)$ phase. [This is marked by ``III'' in panel~(b).] In this case the previously described ``regulation'' process starts immediately, which will decrease the area of the $(D,C,L)$ phase. Finally, when an $L$ domain (marked by green) is at the interface then it will shrink immediately because $M$ is able to utilize the positive impact of network reciprocity. (This process is marked by ``IV'' in the panel.) Altogether, the three elementary processes will reduce the area of the middle zone. As the total area of the three-strategy phase shrinks, it becomes more vulnerable against an external invasion because the local oscillations of $(D,C,L)$ strategies are significant in small patches. (Note that in the middle zone $L$'s would only survive if defectors feed them due to the cyclic dominance.) Consequently, when the width of the middle zone becomes comparable to the typical size of patches then the three-strategy phase can be easily trapped into a homogeneous state. This stage is illustrated in panel~(c) in Fig.~\ref{tolerance_works}. After, independent of which strategy is present at the frontier, the phase becomes an easy prey for the two-strategy $(D,M)$ phase. Finally this solution will invade the whole system (not shown). We stress again that the system will terminate into the same state if we start the evolution from a random state independent of the initial fractions of strategies.

In agreement with the well-mixed case, the application of a moderate tolerance level does not only help strategy $M$ to survive, but it also has  a useful impact on the average payoff of the whole population. This observation is specially important because a previous work highlighted that the introduction of loner strategy is unable to solve the original problem of the public goods game and the average payoff cannot exceed the income of the loner's strategy \cite{hauert_s02}. Therefore, to participate in the joint venture is not an attractive option for loners.

\begin{figure}[ht]
\centerline{\epsfig{file=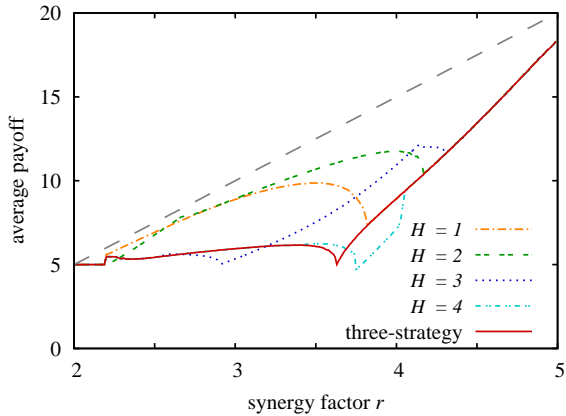,width=8cm}}
\caption{\label{average_payoff}  (Color online) Average payoff in dependence of a synergy factor using different thresholds of tolerance when $G=5$. The cost of inspection is $\gamma=0.04$ for all cases. For comparison, the result of the traditional three-strategy model is also plotted where only pure $D, C$ and $L$ players are present. The highest collective payoff is marked by the dashed grey line which can only be reached in the idealistic case if all players cooperate unconditionally in the group.}
\end{figure}

The concept of tolerance, however, can resolve this dilemma.
Figure~\ref{average_payoff} illustrates the average payoff in dependence of the synergy factor for different threshold values of tolerance at a reasonable cost of inspection. (Note that qualitatively similar behavior can be obtained for higher cost values.) For comparison, the average payoff is also plotted in the traditional model where only the pure ($D,C,L$) strategies are present. In the latter case 
the growth of the general payoff is being hindered by the presence of strategy $L$ no matter how we apply higher $r$. The average income of players can only increase significantly when loners die out. In the latter case, when $r$ is high enough, the network reciprocity can lower the fraction of defectors efficiently which will be followed by the general rise in payoff. 
To evaluate properly the payoff values due to tolerance we have also plotted the highest collective payoff value (marked by the dashed grey line) which can be obtained only if all players cooperate unconditionally in the group. Figure~\ref{average_payoff} shows that the system can be very close to this idealistic state even if $M$ players have to bear an extra cost. In agreement with our previous observation in well-mixed systems, this effect is the strongest at low $r$ values, where cooperation would be unlikely otherwise. 
This feature suggests that the application of tolerance becomes extremely useful when the other cooperator supporting mechanism, based on network reciprocity, becomes fragile.

\begin{figure}[ht]
\centerline{\epsfig{file=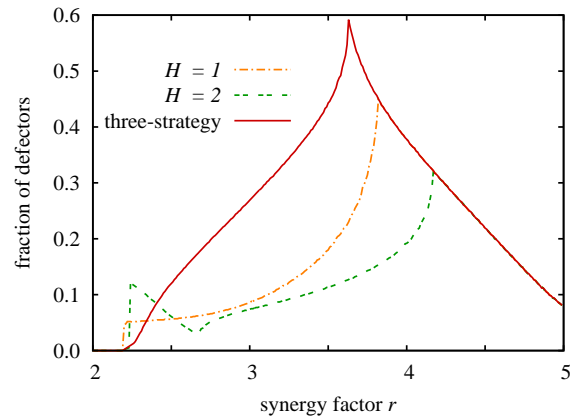,width=8cm}}
\caption{\label{defection}  (Color online) The fraction of defector players in dependence of a synergy factor for those threshold values where the application of tolerance is capable of suppressing the vitality of defectors. For both cases $\gamma=0.04$ was applied as for Fig.~\ref{average_payoff}. For comparison, the fraction of defectors is also plotted when only unconditional $L$ and $C$ strategies fight against defection. Note that defectors could only grow notably when strategy $M$ dies out.}
\end{figure}

Rather counterintuitively, the concept of tolerance of defection is capable of minimizing the occurrence of defectors. This effect is illustrated in Fig.~\ref{defection} where we have plotted the fraction of defectors for the cases when the appropriately chosen tolerance level can result in a notably high average payoff. For comparison, we have also plotted the level of defection in the reference three-strategy $(C,D,L)$ model. As the plot shows, we can reach only limited impact on reducing defection by applying unconditional loners. The latter strategy gives a too ``drastic'' response to defection and cannot utilize the positive effect of network reciprocity. Tolerant players, however, use both sides of the coin: Punish defectors by switching to the loner state if it is inevitable, but remain a cooperator until the last chance. Consequently, they can reach a competitive payoff, which could also be attractive for other players that will reduce the defection level implicitly.

\begin{figure}[ht]
\centerline{\epsfig{file=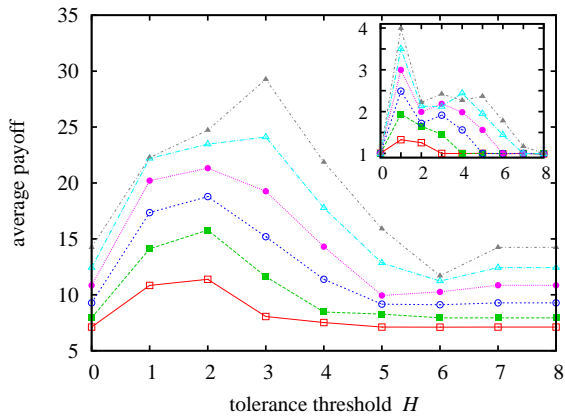,width=8cm}}
\caption{\label{threshold}  (Color online) Average payoff in dependence of the threshold value of tolerance for different synergy factors when a Moore-neighborhood is applied. The inset shows the payoff values for a well-mixed case where the same $G=9$ group size is considered.
For both topologies we used the fixed  $\gamma=0.1$ inspection cost. The applied synergy factors are $r=2.5, 3, 3.5, 4, 4.5$ and $5$ from bottom to top.
Note that $H=0$ corresponds to the three-strategy $(D,C,L)$ model where $M$ players are unable to survive. Because of discrete values of $H$, the dashed lines are just guides to the eye. (The error bars are smaller than the symbol sizes.)}
\end{figure}

The clear advantage of an intermediate tolerance level can be illustrated nicely if we apply larger groups where even more different $H$ threshold values are available. Larger groups can be easily formed if we extend the interaction range from the von Neumann to the Moore neighborhood where players are arranged into $G=9$ group size with their nearest and next-nearest neighbors.
For comparison, in Figure~\ref{threshold} we have also plotted the related payoff values for a well-mixed system where the same group size was applied. These plots demonstrate clearly that having tolerant players in the population is beneficial to the whole society. Furthermore, in a spatial system, where bonds are limited and are maintained for a long time, even a higher tolerance threshold could be the best compromise which provides the highest average payoff for players at some parameter values. 
Naturally, this effect can be even more pronounced for larger group sizes which are typical for human systems \cite{clutton_brock_n09,yang_w_pnas13,szolnoki_pre11c}.

\section{Summary and Conclusion}
\label{summary}

It is our everyday experience that tolerance embraces us whereas the absence of it has serious consequences on the whole community. Our simple model can provide an intuitive explanation for its evolutionary origin: Albeit it might be costly, but it pays to monitor our neighborhood and react on how the cooperation level changes around us. Even if we recognize some defection in our group we should show tolerance towards it because by quitting out from the joint venture we would loose the possible benefit of mutual cooperation. But, of course, we should not be tolerant endlessly because such an attitude takes the system back to the original version of the dilemma where uncontrolled defectors can exploit unconditional cooperators easily. Instead, a delicately adjusted threshold of tolerance can utilize the advantage of both the unconditional cooperator and the loner strategies. Namely, a moderate tolerance threshold helps to utilize the synergy impact of mutual cooperation but it can also keep defection at a bearable level, which altogether can provide a reasonable welfare for the whole community.

It is worth stressing that unconditional cooperator strategy does not necessarily represent a ``naive'' approach from players. There can be those who are generally generous towards others but do not want to invest extra effort to inspect others' acts continuously. Being tolerant, however, involves not only just a forgiving approach towards others but also assumes a permanent monitoring of the neighborhood.

It has been studied intensively how players can avoid being exploited in social dilemmas. One option could be to break adverse ties or leaving an unsatisfactory neighborhood and build new connections on social networks \cite{pacheco_jtb06,van-segbroeck_prl09,lin_yt_pa11,szolnoki_epl09,chen_xj_pre09b,szolnoki_njp09,jiang_ll_pre10,li_y_csf15}. These works focused on the evolving interaction graph and concluded that emerging local homogeneities have a decisive importance on the evolution of cooperation. Indeed, focusing on the similarity of partners or tag-based support is a well-known mechanism, which could provide a clear advantage for cooperation \cite{riolo_n01,traulsen_pre04,masuda_prsb07b,laird_ijbc12}. But some tolerance, according to the present study, might be beneficial, which has crucial importance especially when the average group size in a community is considerably large.

Our paper underlines that the positive impact of tolerance is robust and can be observed both in well-mixed and in structured populations. The effect, however, is more pronounced in a spatial system because network reciprocity augments the basic mechanism. The supporting influence of spatiality could explain the widespread emergence of tolerant behavior \cite{melis_ab06}.

One may claim that strategy $M$ is conceptually similar to a tit-for-tat strategy \cite{hutson_prsb95,rand_jtb09,szolnoki_pre09b,gao_h_csf13}. Indeed, there is some similarity because $M$ players can behave differently in different situations, but the concept of tolerance offers a more sophisticated reaction that is more beneficial to the whole community.
Our last figure illustrates that the best response to varying conditions could be different and sometimes more, sometimes less tolerance provides higher average income, hence a simple reactive strategy would be too rude to respond adequately, especially in the case when multi-point or group interaction is considered \cite{perc_jrsi13}. A logically similar approach could be the possibility of conditional cooperation or conditional participation in joint efforts
\cite{keser_sje00,izquierdo_ss_jtb10,szolnoki_pre12,li_m_pa13}, but the present paper reveals the significant role of additional cost, which was partly ignored in previous works.

Being tolerant to a certain point can also be considered as a threshold game where there is a nonlinear relation between the benefit of the whole group and the proportion of cooperators \cite{heckathorn_asr96,bach_jtb06,pacheco_prsb09,szolnoki_pre10}. Our model, however, is conceptually closer to conditional strategies where the decision is made on a personal level which provides an optimal choice not only for an individual but also for the whole community.  

Lastly, we note that several pioneering works demonstrated the utility of punishment but also highlighted its side effects \cite{panchanathan_n04,hauert_s07}. Namely, it could be effective to control defection, but simultaneously, the usage of punishment may lower the income of both punisher players and those who are punished. On the other hand, reward has a cooperation supporting impact but it also requires an additional source (of reward) to apply it \cite{sigmund_pnas01,cuesta_jtb08,rezaei_ei09,szolnoki_epl10,wu_t_pre12,chen_xj_jtb13}. The presently discussed mechanism, however, offers a simple, but still effective, way on how we can tame defection without losing well-being.

\begin{acknowledgments}
This research was supported by the Hungarian National Research Fund (Grant K-101490) and by the Fundamental Research Funds of the Central Universities of China.
\end{acknowledgments}

\end{document}